\documentclass[%
 preprint,
 amsmath,amssymb,
 aps,
prb,
]{revtex4-2}

\usepackage{graphicx}
\usepackage{dcolumn}
\usepackage{bm}
\usepackage{soul,color}
\usepackage{hyperref}

\begin{document}

\title{Role of interface mixing on coherent heat conduction in periodic and aperiodic superlattices}
\author{Theodore Maranets}
\email{tmaranets@unr.edu}
\author{Evan Doe}
\author{Yan Wang}
\affiliation{Department of Mechanical Engineering, University of Nevada, Reno, Reno, NV, 89557, USA }

\begin{abstract}
Superlattices (SLs) can induce phonon coherence through the periodic layering of two or more materials, enabling tailored thermal transport properties. While most theoretical studies assume atomically sharp, perfect interfaces, real SLs often feature atomic interdiffusion spanning approximately a single atomic layer or more. Such interface mixing can significantly influence phonon coherence and transport behavior. In this study, we employ atomistic wave-packet simulations to systematically investigate the effects of interface mixing on coherent heat conduction. Our analysis identifies two competing mechanisms that govern phonon transport across mixed interfaces: (1) Interface mixing disrupts coherent mode-conversion effects arising from the interface arrangement. (2) The disorder enhances the potential for interference events, generating additional coherent phonon transport pathways. The second mechanism enhances the transmission of Bragg-reflected modes in periodic SLs and most phonons in aperiodic SLs, which otherwise lack coherent mode-conversion in perfect structures. Conversely, the first mechanism dominates in periodic SLs for non-Bragg-reflected modes, where transmission is already high due to substantial mode-conversion. These findings provide insights into the interplay between interface imperfections and phonon coherence.
\end{abstract}

\maketitle

\section{Introduction}

Significant wave interference of phonons in nanoscale thermo-electronic devices can result in coherent phonon transport when the material exhibits secondary periodicity \cite{maldovan2015phonon,volz2016nanophononics,xie2018phonon,zhang2021coherent,anufriev2021review}. Specifically, phonons scattered by periodically arranged features such as interfaces, boundaries, or defects can constructively interfere, giving rise to new vibrational modes resonant with the device’s artificial structure \cite{latour2014microscopic,latour2017distinguishing,maranets2024influence,maranets2024prominent}. These modes, commonly referred to as coherent phonons, exhibit wave-like behavior that contrasts with the particle-like descriptions of phonon transport typically employed in the Boltzmann transport equation and phonon-gas models for heat conduction in structurally simpler solids \cite{simkin2000minimum,ravichandran2014crossover,luckyanova2012coherent,zhu2014phonon,wang2014decomposition,anufriev2016reduction,puurtinen2016low,cui2024spectral}.

The unique wave nature of coherent phonons presents opportunities for manipulation through targeted structural modifications, enabling the tailoring of thermophysical properties \cite{shi2015evaluating}. For instance, semiconductor superlattices (SLs)---composite structures comprising periodically alternating layers of two or more materials---can achieve ultra-low lattice thermal conductivity in the direction perpendicular to the interfaces while maintaining favorable electrical conductivity, a characteristic highly desirable for thermoelectric applications \cite{qian2021phonon}. Structural variations, such as disrupting secondary periodicity through aperiodic layering to form aperiodic SLs, can significantly influence the propagation and characteristics of coherent phonons \cite{wang2015optimization,juntunen2019anderson,chowdhury2020machine,ma2020dimensionality,hu2021direct,maranets2024influence,maranets2024prominent,maranets2024phonon}.

While SLs exhibit neat structures with well aligned interfaces between two materials, the interfaces themselves are seldom perfect. Achieving atomically sharp interfaces remains challenging, if not impossible, even with state-of-the-art thin-film deposition techniques. For instance, molecular beam epitaxy (MBE) of GaAs/AlAs SLs often results in interface mixing at the scale of one or more atomic layers, where GaAs and AlAs interdiffuse by approximately a single atomic layer or more \cite{bode1992interfaces}. Interface mixing becomes even more pronounced with other deposition techniques or material systems, such as Si/Ge SL \cite{fukatsu1992atomistic}. As a common defect in SLs, interface mixing significantly influences coherent phonon transport, scattering, and interference, underscoring the importance of a thorough understanding of its effects \cite{kechrakos1991role,chen1997size,chen1997thermal,sun2010molecular,polanco2015role,luckyanova2018phonon}. Several studies suggest that interface mixing breaks phonon coherence as evidenced by a monotonic reduction in thermal conductivity as period width decreases \cite{daly2002molecular,landry2009effect}. This result contrasts the minimum thermal conductivity observed for periodic SLs with perfect interfaces which indicates a transition between incoherent and coherent phonon dominated heat conduction \cite{capinski1999thermal,venkatasubramanian2000lattice,ravichandran2014crossover,felix2020suppression}. A study by Huberman et al. revealed the lifetimes of coherent phonons belonging to the dispersion relation of the periodic SL are reduced in the case of interface mixing, thus lowering thermal conductivity \cite{huberman2013disruption}. Several other studies also suggest interface mixing disrupts coherent phonon transport by observation of a weaker length-dependence of thermal conductivity in comparison to structures possessing perfect interfaces \cite{wang2015optimization,qiu2015effects,kothari2017phonon,liu2022heat}. Chakraborty et al. showed through molecular dynamics simulations that while interface mixing can scatter coherent phonons, the presence of an intermediate phonon density of states in the mixed regions can increase thermal conductivity in both periodic and aperiodic SLs of short system length by enhancing transport of incoherent phonons across the interfaces \cite{chakraborty2017ultralow}. 

While insightful, inferences about the influence of interface mixing on phonon coherence derived primarily from thermal conductivity data require careful interpretation. Most computational and experimental methodologies are limited in their ability to resolve phonon wave physics and, as a result, struggle to accurately characterize phonon coherence, an inherently wave-based phenomenon \cite{luo2013nanoscale,minnich2015advances,lindsay2019perspective}. Atomistic simulations that model phonons as wave packets offer a powerful approach for studying phonon coherence due to their unique capability to simultaneously define mode, wavevector, and spatial coherence length \cite{schelling2002phonon,tian2010phonon,liang2017phonon,latour2017distinguishing,maranets2023ballistic,maranets2024influence,maranets2024prominent,maranets2024phonon}. These parameters play a critical role in phonon wave interference, from which coherent transport emerges \cite{latour2014microscopic,latour2017distinguishing,maranets2024influence,maranets2024prominent}.

For instance, we recently demonstrated the mode-conversion of incoherent phonons, which follow the bulk dispersion relations of the constituent layer materials, to coherent phonons governed by the dispersion relation of the SL \cite{maranets2024prominent}. While such mode-conversion has been widely hypothesized for periodic SLs, direct proof has been lacking. In the case of aperiodic SLs, coherent mode-conversion facilitates significant phonon transmission, which helps to explain the non-trivial thermal conductivities observed in these structures. Additionally, we have employed the wave-packet method to examine coherent phonon behaviors influencing the temperature and length dependencies of thermal conductivity in aperiodic SLs \cite{maranets2024influence,maranets2024phonon}.

Building upon our prior work, here in this work, we aim to apply atomistic wave-packet simulations to investigate the impact of interface mixing on coherent heat conduction. This approach provides valuable insights into the interplay between structural modifications and phonon coherence, enhancing our understanding of heat transport in complex material systems.

\section{Methodology}
\subsection{Material system\label{sec:material_system}}

Our model material system consists of two Lennard-Jones face-centered cubic (FCC) crystals with lattice parameters equivalent to those of solid argon. The two materials, referred to as m40 and m90, have molecular weights of 40 g mol$^{-1}$ and 90 g mol$^{-1}$, respectively. Both materials share identical force constants, resulting in no elastic contrast within the system---only a mass contrast. The Lennard-Jones potential well depth is $\varepsilon = 0.1664$ eV, with a zero-crossing distance of $\sigma = 3.4$ \AA. A cutoff radius of $2.5\sigma$ is used. The well depth, which is 16 times that of solid argon, is chosen to approximate covalently bonded semiconductor systems, such as Si/Ge and GaAs/AlAs \cite{landry2008complex,huberman2013disruption,wang2014decomposition,wang2015optimization,chakraborty2017ultralow,chakraborty2020complex,maranets2023lattice}. The relaxed lattice constant, corresponding to the size of one unit cell, is $1 \, \text{UC} = 5.27$ \AA.  

The periodic SL is composed of alternating layers of m40 and m90, each 4 UC thick, resulting in a period width of 8 UC. The aperiodic SL is derived from the periodic SL using the methodology outlined in Ref. \cite{wang2015optimization}. As demonstrated in our previous work, this algorithm generates sufficiently disordered aperiodic layer thicknesses to suppress coherent phonon transport in SLs \cite{chakrabortyACSAMI2020}. To model interface mixing, we introduce a mixing fraction $f$, randomly replacing $f\%$ of atoms in the interfacial layer, drawn from a uniform distribution, with atoms of the other species. The case of perfect interfaces corresponds to $f = 0\%$. A large cross-sectional area of $8 \times 8$ UC is employed to ensure representative mixed interfaces.

\subsection{Atomistic phonon wave-packet simulation}

\begin{figure}
    \centering
    \includegraphics[width=0.9\textwidth]{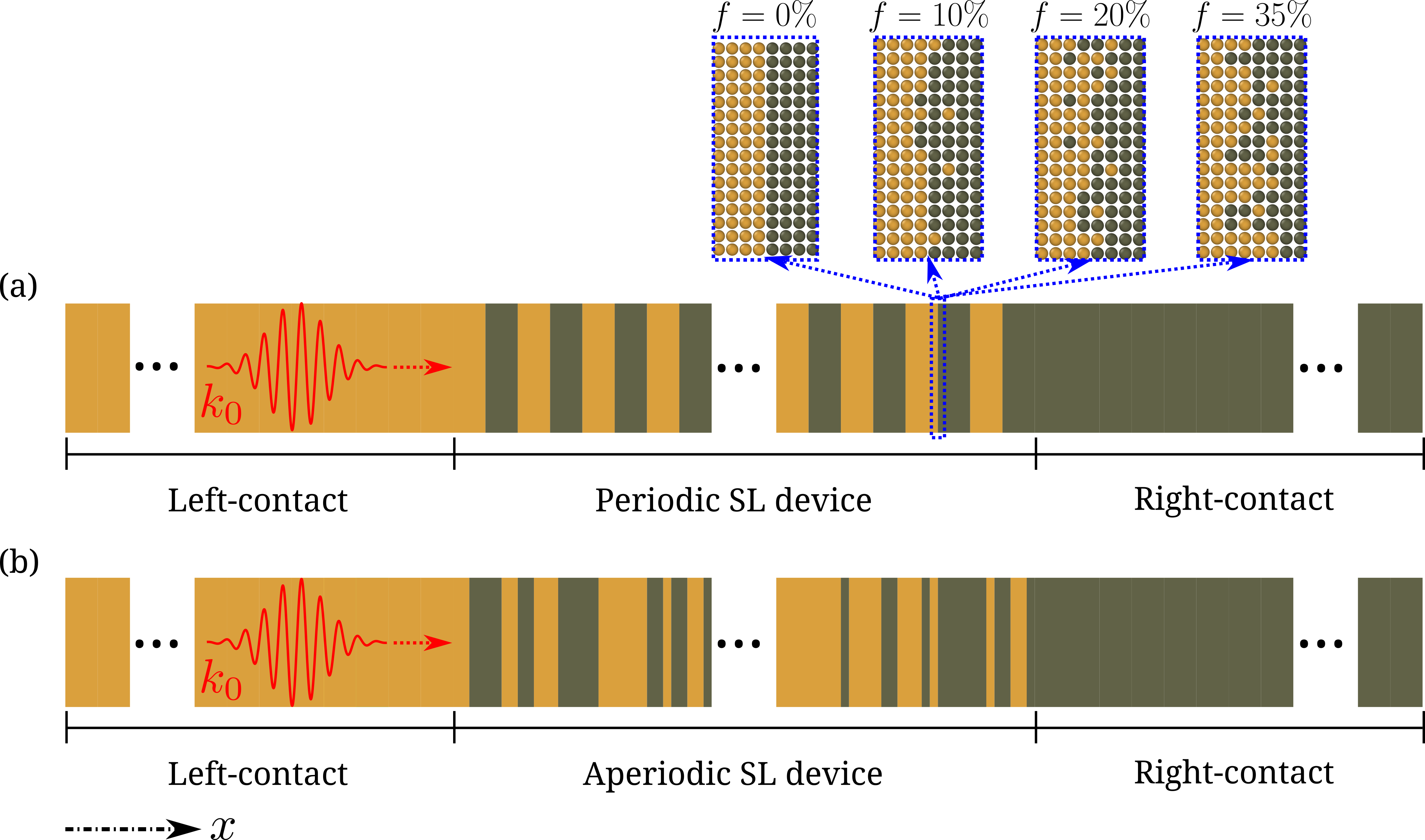}
    \caption{Schematic illustrations of the simulation domains for periodic SL (a) and aperiodic SL (b) devices; visualizing the wave-packet simulation process and the interface mixing fraction $f$. The incident LA-mode incoherent phonon wave-packet centered at wavevector $k_{0}$ in reciprocal-space is generated in the left-contact and allowed to travel into the SL device. The total energies of the left-contact, right-contact, and device are recorded over the simulation duration to compute the transmission across the device per Eqn.~\ref{eqn:transmission_calc}. The spatial coherence length (real-space width of the wave-packet) is set to four times the device length to accentuate interference and consequently mode-conversion effects as elucidated in our prior study \cite{maranets2024influence}. The device length is 64 periods while the size of each contact is 608 periods. The sizes of the illustrated wavelength and spatial coherence length are not to scale. Atomic visualization of the interfaces is done with the OVITO software \cite{ovito}.}
    \label{fig:system_diagram}
\end{figure}

The wave-packet method applied in this study is based off the approach by Schelling et al. \cite{schelling2002phonon} We simulate the propagation of a longitudinal-acoustic (LA) incoherent phonon wave-packet into a SL device and compute the transmission across the device. Schematics and details of the simulation domains are presented in Fig.~\ref{fig:system_diagram}. We explore the impact of mixing fraction $f$ on the transmission spectra of periodic and aperiodic SLs.

The wave-packet is generated at the beginning of the simulation in the left-contact by setting the displacement and velocity of the $i$th atom in the $n$th unit cell according to the following equations:
\begin{equation}
u_{i,n} = \frac{A_{i}}{\sqrt{m_{i}}}\varepsilon_{k_{0},i}\exp{(i[k_{0}\cdot(x_{n} - x_{0})-\omega_{0}t]})\exp{(-4(x_{n}-x_{0}-v_{g0}t)^{2}/l_{c}^{2})},\qquad 
\label{eqn:pwp_disp}
\end{equation}
\begin{equation}
v_{i,n} = \frac{\partial u_{i,n}}{\partial t},\qquad
\label{eqn:pwp_vel}
\end{equation}
where $\varepsilon_{k_{0},i}$, $\omega_{0}$, and $v_{g0}$ are the eigenvector, frequency, and group velocity values associated with wavevector $k_{0}$ from the bulk dispersion relation of the m40 conceptual material defined in Section~\ref{sec:material_system}. $m_{i}$ is the mass of atom $i$ and $x_{n}$ is the position of the $n$th unit cell along the direction of wave-packet propagation. The wave amplitude $A_{i}$, the initial wave-packet position $x_{0}$, and the spatial coherence length $l_{c}$ are user-specified values. The real parts of Eqns.~\ref{eqn:pwp_disp} and \ref{eqn:pwp_vel} evaluated at time $t=0$ are the perturbation values used to generate the wave-packet. Periodic boundary conditions are set in all three dimensions. The simulations are run in the micro-canonical ensemble with the LAMMPS software \cite{thompson2022lammps}. 

The transmission of a wave-packet across the periodic and aperiodic SL devices is computed with the following equation:
\begin{equation}
    \mathcal{T}=\frac{E_{rc,final}}{E_{lc,initial}},\qquad
    \label{eqn:transmission_calc}
\end{equation}
where $E_{rc,final}$ and $E_{lc,initial}$ are the final and initial total atomic energies in the right and left-contacts, respectively. The simulation duration is sufficiently long such that all the incident wave-packet energy has been either reflected or transmitted through the device by the end of the simulation. $x_{0}$ is also set so that the no wave-packet energy exists within the device at the beginning of the simulation.

\subsection{Equilibrium Molecular Dynamics}

We perform equilibrium molecular dynamics (EMD) simulations using the LAMMPS software \cite{thompson2022lammps} to calculate the thermal conductivity perpendicular to the interfaces of periodic and aperiodic SL supercells as a function of the interface mixing fraction \(f\). The supercell consists of 8 periods along the cross-plane direction with a cross-sectional area of \(8 \times 8\) unit cells. Periodic boundary conditions are applied in all three dimensions.

The simulations are conducted with a timestep of 1 fs. The structure is initially relaxed at 30~K and 1~bar in the isothermal–isobaric ensemble for 1.5~ns. Following this, the system is evolved in the micro-canonical ensemble for 30~ns, during which the heat fluxes are recorded \cite{surblys2019application}. The thermal conductivity \(\kappa\) is then calculated using the Green-Kubo formalism \cite{green1954markoff,kubo1957statistical,mcgaughey2006phonon} to evaluate the $\kappa$ along the cross-plane direction, specifically, 
\begin{equation}
\kappa_{x} = \frac{V}{k_{B}T^2}\int_{0}^{\infty}<J_{x}(t)J_{x}(0)>\,dt ,
    \label{eqn:G-K}
\end{equation}
where $k_{B} $ is the Boltzmann constant, $T$ is the ambient temperature, $V$ is the system volume, $t$ is the time, and $J_{x}$ is the heat flux in the $x$ direction. $<J_{x}(t)J_{x}(0)>$ is the heat flux autocorrelation function. The $\kappa$ values reported in this paper are computed from the average of four Green-Kubo curves taken from four EMD simulations with different velocity initializations.
 
The full 64 period structures used in our wave-packet simulations are not employed here because EMD simulations based on the Green-Kubo approach require an extensive number of simulation steps to achieve convergence, particularly for materials with low thermal conductivity, such as aperiodic SLs. Based on Ref.~\cite{wang2015optimization}, an 8-period configuration is sufficient to accurately predict the cross-plane thermal conductivity of both periodic and aperiodic SLs. This choice is justified as follows. For periodic SLs, the application of periodic boundary conditions along the cross-plane direction effectively models an infinite number of periods. For aperiodic SLs, our prior study demonstrated that 8 periods are adequate to induce coherent mode-conversions characteristic of many-period structures \cite{maranets2024phonon}. Therefore, we use 8-period SLs in our EMD simulations to optimize computational efficiency while maintaining an acceptable level of accuracy.

\section{Results and Discussion}

\subsection{Transmission: Periodic SL}

\begin{figure}
    \centering
    \includegraphics[width=0.9\textwidth]{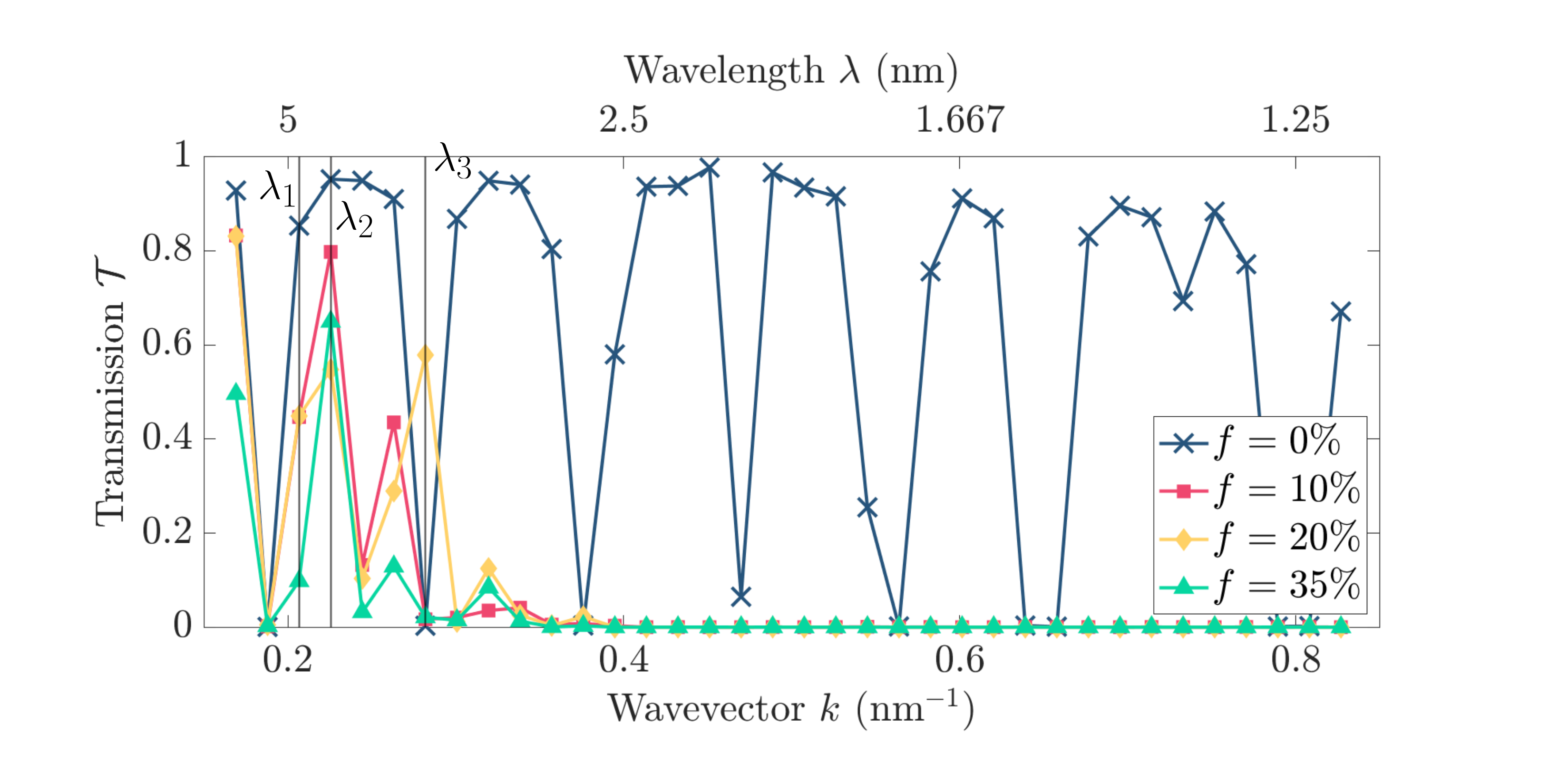}
    \caption{Transmission $\mathcal{T}$ versus wavevector $k$ (or inverse of wavelength $\lambda$) for the LA-mode incoherent phonon wave-packet propagating through periodic SL devices of differing mixing fractions $f$ as shown in Fig.~\ref{fig:system_diagram}a. $\mathcal{T}$ is evaluated with Eqn.~\ref{eqn:transmission_calc}. Key wavelengths $\lambda_{1} = 4.84$ nm, $\lambda_{2} = 4.43$ nm, and $\lambda_{3}=3.55$ nm analyzed in this paper are indicated by the vertical black lines in the figure. }
    \label{fig:transmission_SL}
\end{figure}

\begin{figure}
    \centering
    \includegraphics[width=0.55\textwidth]{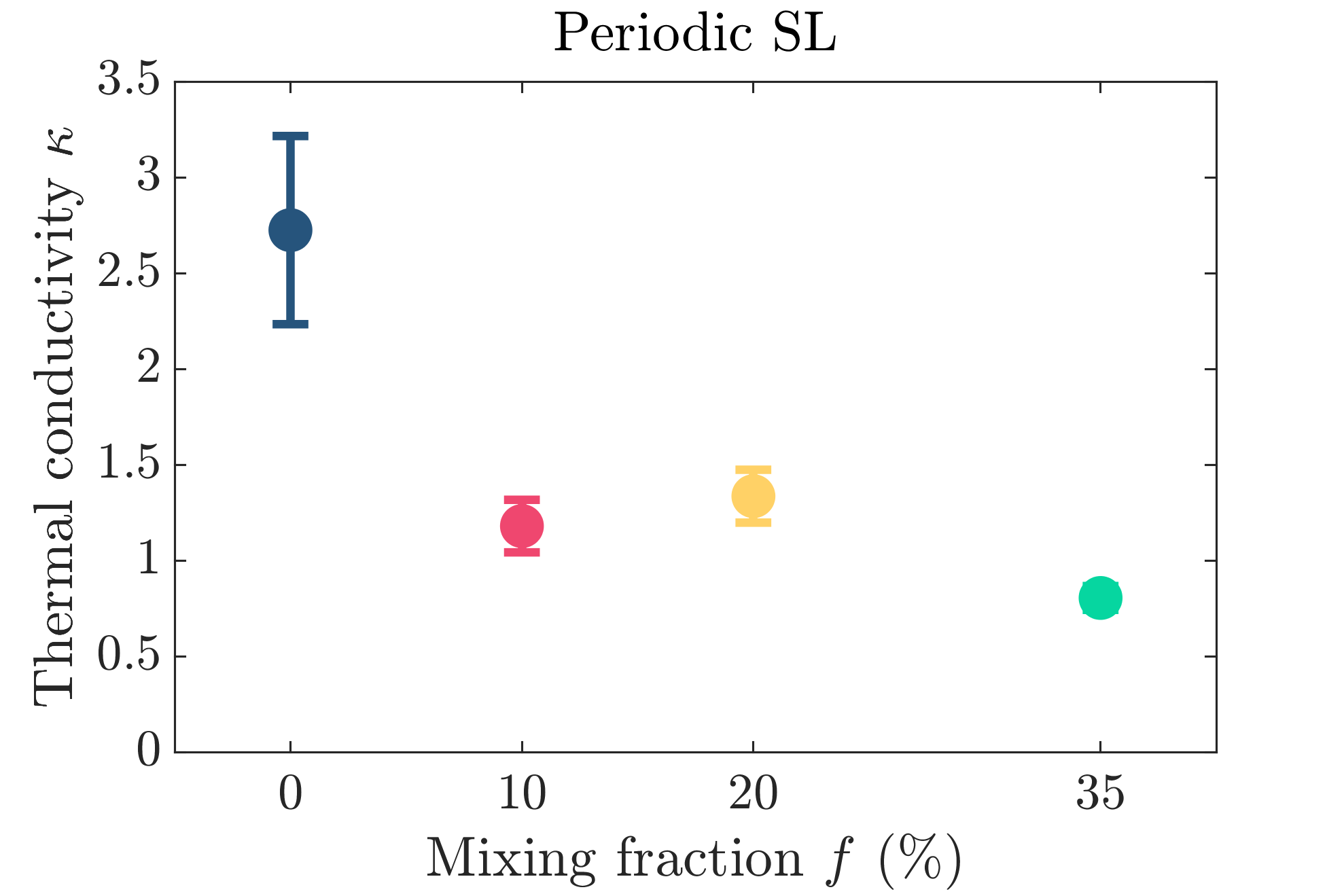}
    \caption{Cross-plane thermal conductivity $\kappa$ (at 30 K) versus mixing fraction $f$ for 8 period periodic SL supercells computed from EMD simulations. $\kappa$ is evaluated with Eqn.~\ref{eqn:G-K}.}
    \label{fig:kappa_SL}
\end{figure}

We begin with the transmission spectra of the periodic SL, examining the impact of mixing fractions of $f=10\%$, $f=20\%$, and $f=35\%$ on energy transport. The results are presented in Fig.~\ref{fig:transmission_SL}. First, we analyze the spectrum of the periodic SL with perfect interfaces ($f=0\%$) and observe an oscillatory pattern, consistent with numerous prior studies on similar structures \cite{hurley1987imaging,tamura1988acoustic,tamura1988acousticmulti,tamura1989localized,tanaka1998phonon,maranets2024prominent,maranets2024phonon}. The high transmission values are attributed to coherent mode-conversion \cite{schelling2003multiscale,jiang2021total,maranets2024prominent} while the transmission dips result from destructive Bragg interference \cite{hurley1987imaging,tamura1988acoustic,tamura1988acousticmulti,tamura1989localized,tanaka1998phonon,maranets2024prominent}.

Moving on to the periodic SLs with mixed interfaces, we observe a reduction in transmission for low wavevector phonons and a striking complete suppression of transmission at higher wavevectors. Furthermore, for low-wavevector phonons, the mixing fraction significantly influences the transmission magnitude $\mathcal{T}$, which generally decreases as the mixing fraction $f$ increases. However, a monotonic relationship between $\mathcal{T}$ and $f$ is not always observed. For instance, at the wavelength $\lambda_{3}$ (Fig.~\ref{fig:transmission_SL}), the transmission exhibits non-monotonic behavior. As illustrated by wavelet transform analysis later in this paper, this is due to the complex dependence of mode-conversion properties on wavelength and interface structure, which does not necessarily follow a simple monotonic trend with $f$ or other parameters.

Regarding the zero transmission of high wavevector phonons shown in Fig.~\ref{fig:transmission_SL}, our previous work demonstrated that the absence of transmission in periodic and aperiodic SLs stems from the inability of the incident incoherent phonon wave-packet to mode-convert into a coherent phonon \cite{maranets2024prominent}. In the present study, this inability is attributed to the interface mixing present at each individual interface. The pronounced impact of interface mixing on phonon transmission at high wavevectors (shorter wavelengths) is consistent with prior studies, which have shown that high wavevector phonons exhibit greater sensitivity to the character of interfaces \cite{huberman2013disruption,qiu2015effects,kothari2017phonon,xie2022impacts,maranets2024influence,maranets2024prominent,maranets2024phonon} and, more generally, other types of defects such as impurities, roughness, etc. 

For low wavevector phonons, whose wavelengths are significantly larger than the dimensions of the mixed interface region, the impact of the mixing fraction is much less than the case of high wavevectors, as shown in Fig.~\ref{fig:transmission_SL}. Nevertheless, the general reduction in transmission with increasing mixing fraction aligns with prior studies suggesting that mixed interfaces scatter coherent phonons, thereby weakening energy transport \cite{yang2003partially,huberman2013disruption,wang2015optimization}. Our EMD simulations of periodic SL supercells (Fig.~\ref{fig:kappa_SL}) further support this argument, demonstrating much lower cross-plane thermal conductivity of SLs with mixed interfaces than that of the perfect SL. 

\subsection{Transmission: Aperiodic SL}

\begin{figure}
    \centering
    \includegraphics[width=0.9\textwidth]{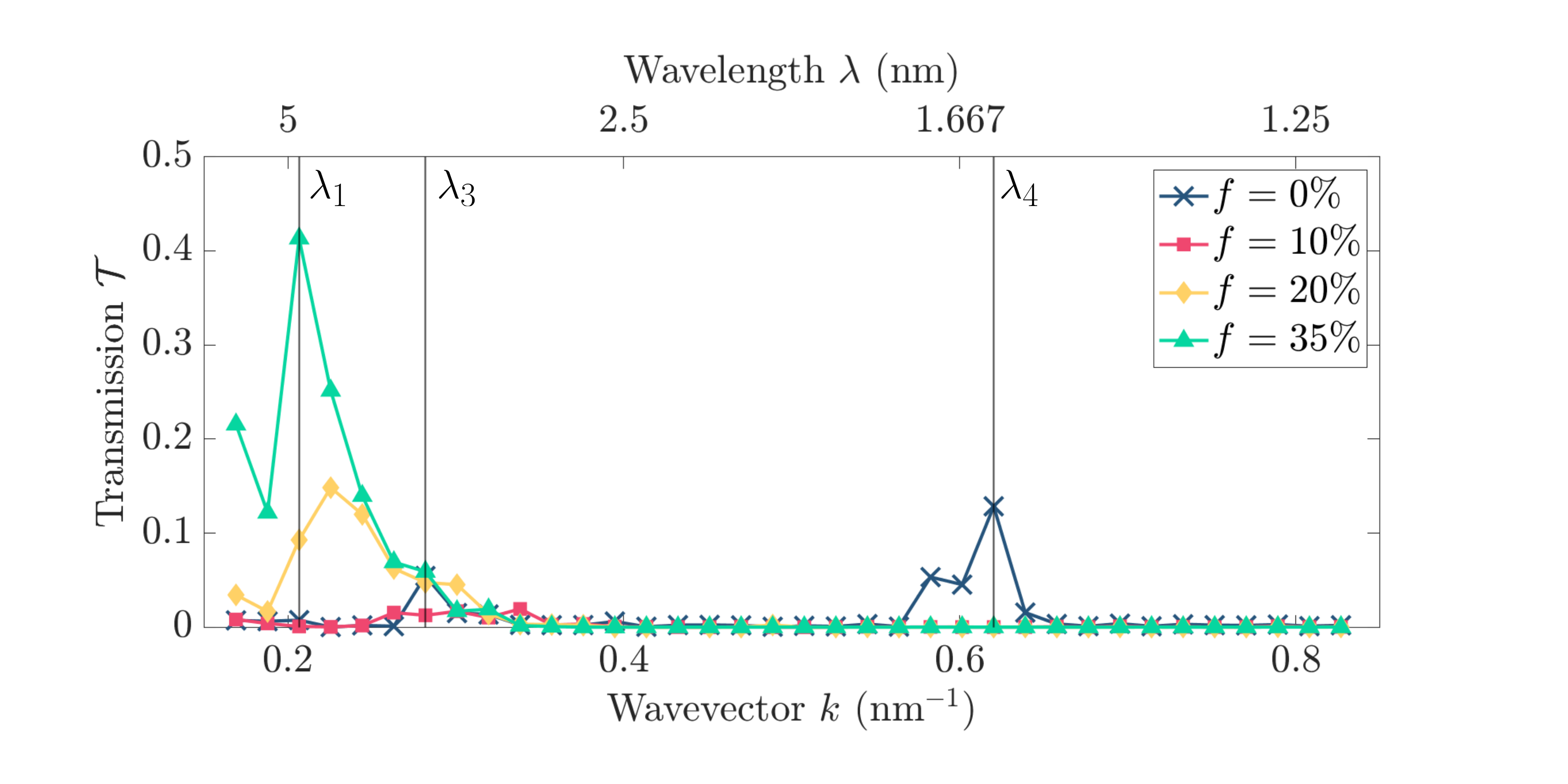}
    \caption{Transmission $\mathcal{T}$ versus wavevector $k$ (or inverse of wavelength $\lambda$) for the LA-mode incoherent phonon wave-packet propagating through aperiodic SL devices of differing mixing fractions $f$ as shown in Fig.~\ref{fig:system_diagram}b. $\mathcal{T}$ is evaluated with Eqn.~\ref{eqn:transmission_calc}. Key wavelengths $\lambda_{1} = 4.84$ nm, $\lambda_{3}=3.55$ nm, and $\lambda_{4}=1.61$ nm analyzed in this paper are indicated by the vertical black lines in the figure. }
    \label{fig:transmission_RML}
\end{figure}

\begin{figure}
    \centering
    \includegraphics[width=0.55\textwidth]{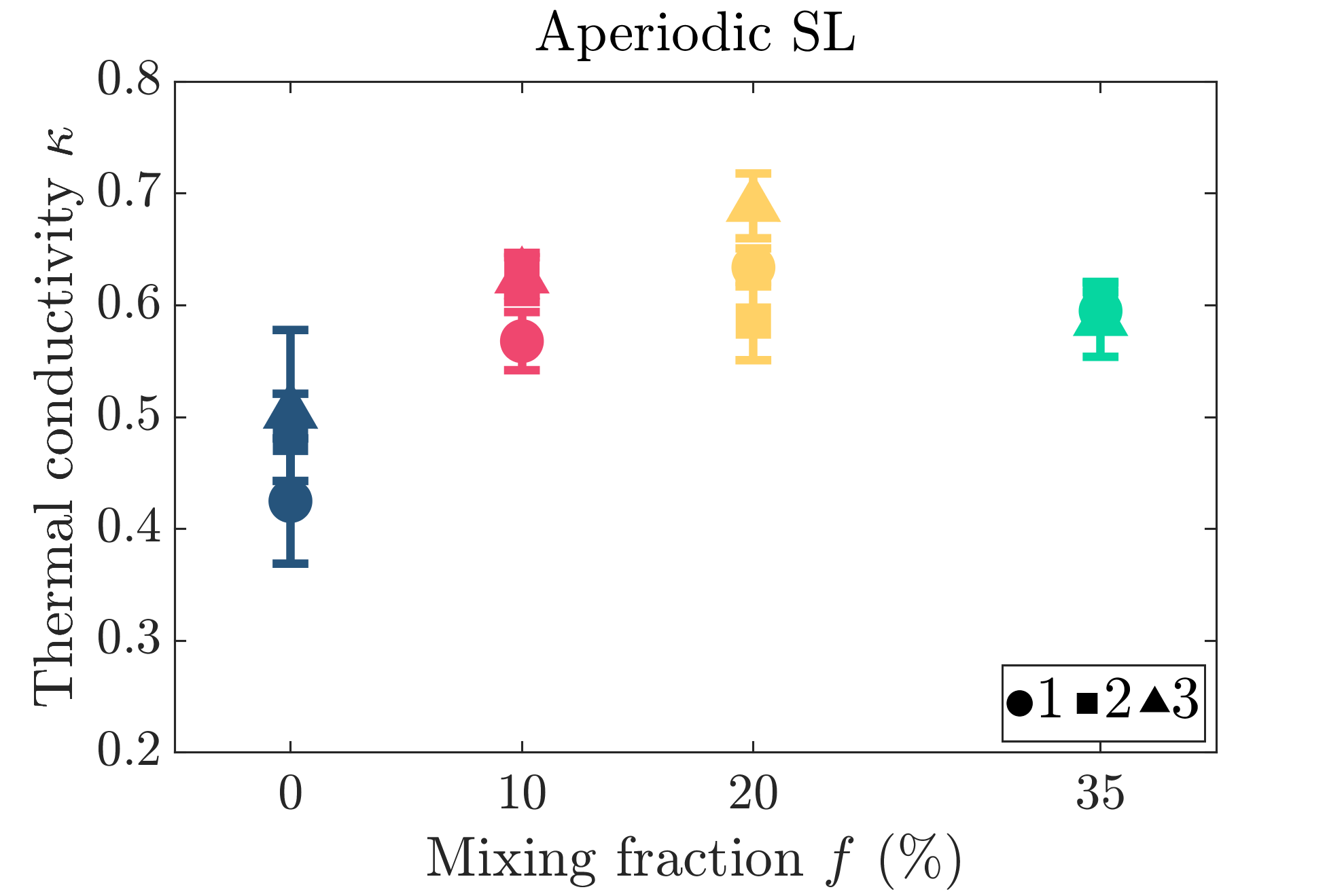}
    \caption{Cross-plane thermal conductivity $\kappa$ versus mixing fraction $f$ for 8 period aperiodic SL supercells computed from EMD simulations. $\kappa$ is evaluated with Eqn.~\ref{eqn:G-K}. The circle, square, and triangle data points delineate the values for three different aperiodic patterns.}
    \label{fig:kappa_RML}
\end{figure}

The impact of interface mixing on phonon behaviors in aperiodic SLs is less understood. Fig.~\ref{fig:transmission_RML} displays the transmission spectra of aperiodic SLs with mixing fractions of $f=10\%$, $f=20\%$, and $f=35\%$, along with the structure with perfect interfaces ($f=0\%$). The transmission spectra for the perfect aperiodic SL closely align with the findings reported in our previous wave-packet investigations \cite{maranets2024prominent,maranets2024phonon}, particularly demonstrating zero transmission across most wavevectors except for a prominent peak corresponding to coherent mode-conversion.

In contrast to the periodic SL, which exhibits a reduction in the transmission of low-wavevector phonons with increasing mixing fraction, the aperiodic SL shows an enhancement in transmission at low wavevectors as the mixing fraction increases. This enhanced transmission is further reflected in the thermal conductivity trends shown in Fig.~\ref{fig:kappa_RML}, which exhibit an overall increasing trend with mixing fraction. This behavior is in stark contrast to the periodic SL, whose thermal conductivity is substantially reduced by interface mixing, as shown in Fig.~\ref{fig:kappa_SL}. At higher wavevectors, where the perfect aperiodic SL exhibits a prominent transmission peak at $\lambda_{4}$ (Fig.~\ref{fig:transmission_RML}), the structures with mixed interfaces display complete suppression of transmission, similar to the behavior observed in periodic SLs (Fig.~\ref{fig:transmission_SL}).

Enhanced heat conduction due to interface mixing in aperiodic SLs has been captured in several prior studies. Notably, English et al. revealed through non-equilibrium molecular dynamics that phonon transmission across a single interface can be increased through interface mixing \cite{english2012enhancing}. A later study by Tian et al. employing the atomistic Green's function method also observed this effect \cite{tian2012enhancing}.
The observed increase in transmission can be attributed to the mixed interface possessing an intermediate vibrational density of states (vDOS) that lies between the vDOS of the two constituent materials forming the interface. This intermediate vDOS effectively acts as a bridge, facilitating heat transfer across the two regions with distinct vDOS---a concept central to the phonon-bridge theory. The impact of interface mixing on the thermal conductivity of aperiodic SLs has been explored in subsequent studies by Qiu et al. \cite{qiu2015effects} and Chakraborty et al. \cite{chakraborty2017ultralow}. Both investigations demonstrated that aperiodic SLs with mixed interfaces exhibit higher thermal conductivity compared to their counterparts with perfect interfaces. Specifically, Chakraborty et al. revealed through non-equilibrium molecular dynamics simulations that the mixed interfacial regions of the aperiodic SL exhibit an intermediate vDOS, supporting the phonon-bridge mechanism \cite{chakraborty2017ultralow}. Additionally, Qiu et al. hypothesized that the three-dimensional spatial disorder introduced by interface mixing may be less effective in disrupting coherent phonon transport via localization compared to the one-dimensional disorder present in the perfect aperiodic SL \cite{qiu2015effects}.

However, the enhancement of energy transport through these mechanisms is not universally observed. Ni et al. found that interface mixing and aperiodic layering act as additive factors in reducing the thermal conductivity of graphene/h-BN SLs \cite{ni2024suppressing}. Specifically, both structural modifications scatter and localize coherent phonons, leading to decreased transmission. A recent study by Cui et al., however, revealed that graphene/h-BN SLs are a unique material system where the two constituent materials possess comparable heat flux spectra \cite{cui2025spectral}. Consequently, interface mixing in graphene/h-BN SLs behaves analogously to alloying, where impurity scattering suppresses phonon transport. In contrast, the constituent materials of the SL studied here, m40 and m90, exhibit significantly different vDOS and heat flux spectra \cite{chakraborty2017ultralow,cui2025spectral}. This suggests that the observed increase in energy transport in the mixed aperiodic SL can be attributed to the formation of intermediate vDOS and/or the weakening of localization pathways. Nevertheless, the underlying mechanisms responsible for enhanced transmission at long wavelengths and suppressed transmission at shorter wavelengths, where the perfect structure exhibits prominent transmission, remain unclear.

\subsection{Reciprocal-space wavelet transforms}

\begin{figure}
    \centering
    \includegraphics[width=\textwidth]{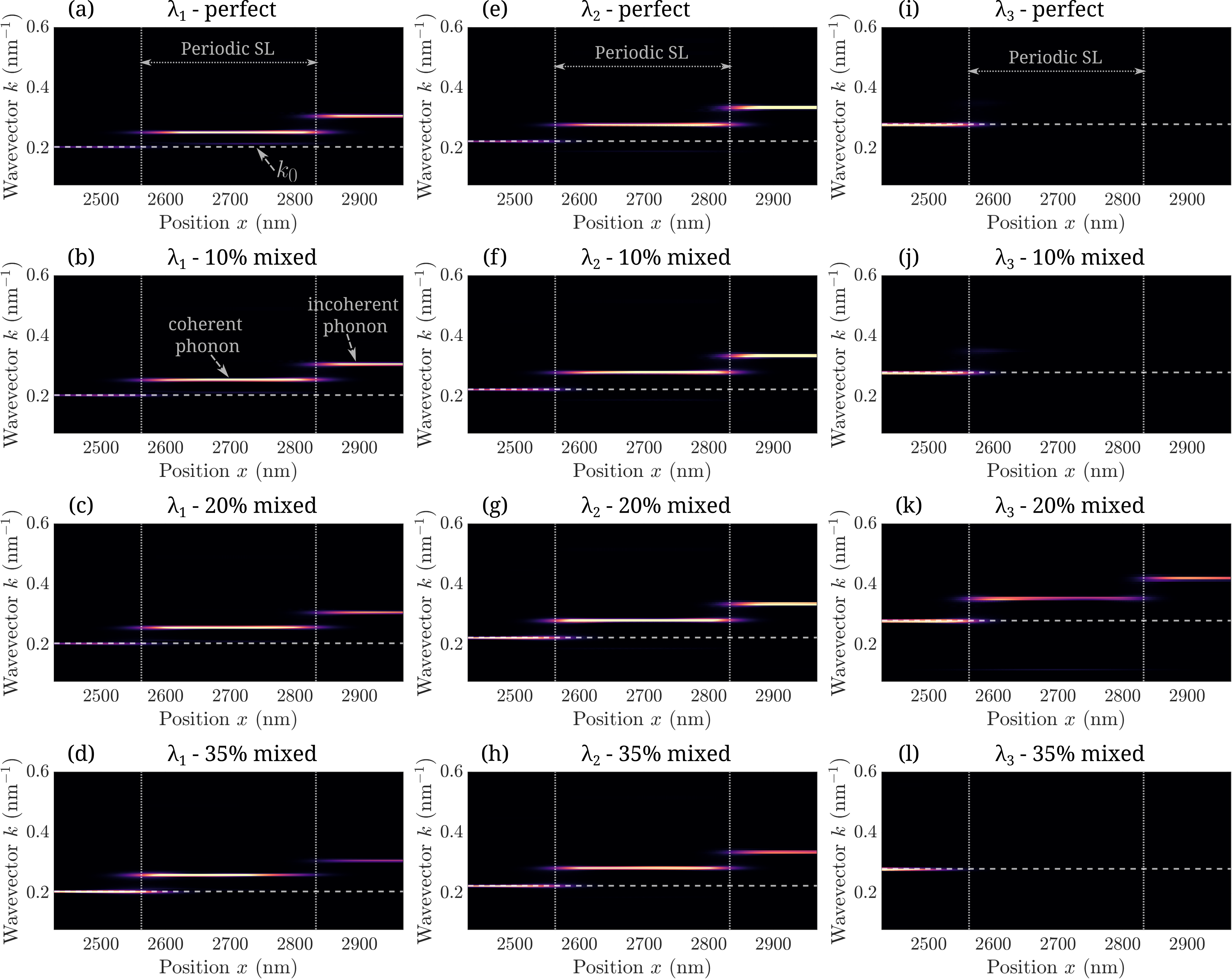}
    \caption{(a-l): Snapshots of the reciprocal-space wavelet transform when the incoherent phonon wave-packet propagates through the periodic SL devices of differing mixing fractions for incident wavelengths $\lambda_{1} = 4.84$ nm, $\lambda_{2} = 4.43$ nm, and $\lambda_{3}=3.55$ nm. The dotted vertical lines denote the position and length of the periodic SL device. The dashed horizontal line denotes the central wavevector $k_{0}$ of the incident phonon. Illuminated regions in the heat map correspond to mode excitations. Consequently, excitations outside the $k_{0}$ line indicate coherent mode-conversion as revealed in Ref.~\cite{maranets2024prominent}.}
    \label{fig:wavelet_SL}
\end{figure}

To study the mechanisms affecting phonon transmission in mixed periodic and aperiodic SLs, we analyze the phonon behaviors taking place in the wave-packet simulation with the wavelet transform. With this tool, the dynamics of phonon wave transport can be probed in real and reciprocal-spaces simultaneously, making it useful for identifying coherent mode-conversion \cite{baker2012application,maranets2024influence,maranets2024prominent,maranets2024phonon}.

Fig.~\ref{fig:wavelet_SL} displays the reciprocal-space wavelet transforms for several wavelengths marked in Fig.~\ref{fig:transmission_SL}, which corroborate our findings for the periodic SL outlined in the previous section. Specifically, transmission in the periodic SL is facilitated by coherent mode-conversion, and factors affecting such conversion accordingly modify transmission. Figs.~\ref{fig:wavelet_SL}a-\ref{fig:wavelet_SL}h show that for $\lambda_{1}=4.84$ nm and $\lambda_{2}=4.43$ nm, non-zero transmission corresponds to coherent mode-conversion in the periodic SL, regardless of the presence of mixed interfaces. In contrast, total suppression of transmission is evident from the absence of mode-conversion, as seen in Figs.~\ref{fig:wavelet_SL}i,~\ref{fig:wavelet_SL}j, and~\ref{fig:wavelet_SL}l. Interestingly, the periodic SL with $f=20\%$ accentuates the mode-conversion of $\lambda_{3}=3.55$ nm (Fig.~\ref{fig:wavelet_SL}k), which is otherwise Bragg reflected. Consequently, as shown in Fig.~\ref{fig:transmission_SL}, the transmission coefficient for the $f=20\%$ SL is nearly 60\% at $\lambda_{3}$, while it is almost zero for periodic SLs with $f=0\%$, $f=10\%$ or $f=35\%$. This further confirms the critical role of coherent mode-conversion in determining the transmission of incoherent phonons across SL structures.

\begin{figure}
    \centering
    \includegraphics[width=\textwidth]{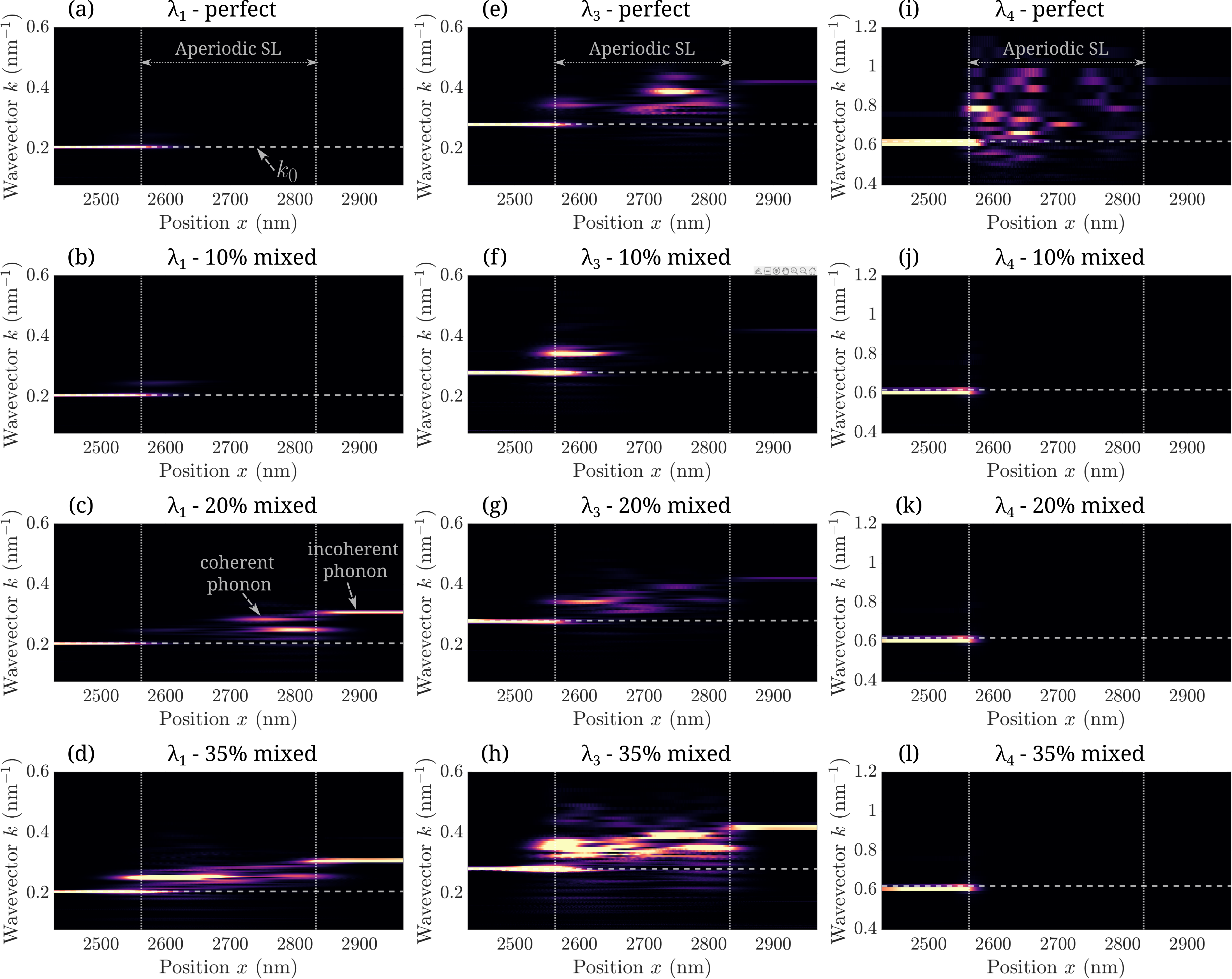}
    \caption{(a-l): Snapshots of the reciprocal-space wavelet transform when the incoherent phonon wave-packet propagates through the aperiodic SL devices of differing mixing fractions for incident wavelengths $\lambda_{1} = 4.84$ nm, $\lambda_{3}=3.55$ nm, and $\lambda_{4}=1.61$ nm. The dotted vertical lines denote the position and length of the aperiodic SL device. The dashed horizontal line denotes the central wavevector $k_{0}$ of the incident phonon. Illuminated regions in the heat map correspond to mode excitations. Consequently, excitations outside the $k_{0}$ line indicate coherent mode-conversion as revealed in Ref.~\cite{maranets2024prominent}.}
    \label{fig:wavelet_RML}
\end{figure}

Regarding the aperiodic SL, we analyze the reciprocal-space wavelet transforms, presented in Fig.~\ref{fig:wavelet_RML}, for several wavelengths of interest marked in Fig.~\ref{fig:transmission_RML}. Beginning with the long-wavelength phonon $\lambda_{1} = 4.84~\text{nm}$, which shows increasing transmission with mixing fraction, Figs.~\ref{fig:wavelet_RML}a--\ref{fig:wavelet_RML}d reveal that this enhanced transmission in the mixed aperiodic SLs arises from accentuated mode-conversion. This observation aligns with the established understanding that the magnitude of transmission in a SL is determined by the extent of mode-conversion \cite{maranets2024prominent,maranets2024phonon}. Similarly, wavelet transforms for the $\lambda_{3} = 3.55~\text{nm}$ phonon, shown in Figs.~\ref{fig:wavelet_RML}e--\ref{fig:wavelet_RML}h, support this notion. Specifically, $\lambda_{3}$ exhibits significantly less mode-conversion in the $10\%$ mixed aperiodic SL (Fig.~\ref{fig:wavelet_RML}f) compared to the other structures (Figs.~\ref{fig:wavelet_RML}e,~\ref{fig:wavelet_RML}g, and~\ref{fig:wavelet_RML}h). This reduced mode-conversion is reflected in the transmission spectra (Fig.~\ref{fig:transmission_RML}), where $\lambda_{3}$ shows the weakest transmission in the $10\%$ mixed aperiodic SL. For $\lambda_{4}=1.61$ nm, which exhibits prominent transmission in the perfect aperiodic SL but not in the mixed structures, Figs.~\ref{fig:wavelet_RML}j--\ref{fig:wavelet_RML}l reveal that the suppression of transmission arises from the absence of mode-conversion, akin to the behavior observed in the periodic SL.

\subsection{Discussion}

The connection between transmission and coherent mode-conversion as well as the factors influencing mode-conversion explain the observed transmission behaviors in the periodic and aperiodic SLs. Interface mixing has two competing effects on phonon transport, as illustrated below. 

First, interface mixing scatters incoherent phonons, disrupting their ability to undergo mode-conversion due to constructive interference associated with the arrangement of interfaces. The attenuation of coherent mode-conversion leads to reduced transmission, as observed for long-wavelength phonons in the periodic SL (Fig.~\ref{fig:transmission_SL}), $\lambda_{3}$ in the $10\%$ mixed aperiodic SL (Fig.~\ref{fig:transmission_RML}), $\lambda_{4}$ in all mixed aperiodic SLs (Fig.~\ref{fig:transmission_RML}), and short-wavelength phonons in both mixed periodic and aperiodic SLs (Figs.~\ref{fig:transmission_SL} and \ref{fig:transmission_RML}). As demonstrated in our prior work \cite{maranets2024prominent}, the broadly high transmission of incoherent phonons in periodic SLs results from significant coherent mode-conversion. Consequently, scattering by mixed interfaces, which degrades conditions for mode-conversion, substantially hinders transmission. Furthermore, the prominent transmission peak in the aperiodic SL (e.g., the $\lambda_{4}$ mode in Fig.~\ref{fig:transmission_RML}), where the majority of the transmission spectrum is near-zero, is attributed to coherent mode-conversion enabled by constructive interference from the specific layer thickness pattern of the aperiodic SL. Interface mixing disrupts this condition, thereby eliminating the prominent peak, as seen in the red, yellow, and green data points at $\lambda_{4}$ in Fig.~\ref{fig:transmission_RML}.

Second, scattering at mixed interfaces introduces new opportunities for interference, enhancing transmission in specific cases. For instance, the enhancement of long-wavelength phonon transmission in mixed aperiodic SLs (e.g., Fig.~\ref{fig:transmission_RML}) through accentuated mode-conversion (Figs.~\ref{fig:wavelet_RML}c and \ref{fig:wavelet_RML}d) seems to contradict the first effect. Instead, this enhancement can be attributed to the constructive interference introduced by interface scattering. This behavior is observed for $\lambda_{1}$ and partially for $\lambda_{3}$ in the aperiodic SL (Figs.~\ref{fig:wavelet_RML}a--\ref{fig:wavelet_RML}h) as well as $\lambda_{3}$ in the $10\%$ mixed periodic SL (Fig.~\ref{fig:wavelet_SL}j), where $\lambda_{3}$ is a mode Bragg-reflected in the perfect structure.

In general, a higher degree of interface mixing enhances the potential for mode-conversion and improves the intermediate vDOS alignment with the vDOS of the adjacent layers, thereby amplifying the phonon bridging effect. This phenomenon accounts for the observed increase in the transmission of the $\lambda_{1}$ and $\lambda_{3}$ modes in aperiodic SLs (Fig.~\ref{fig:transmission_RML}) as the mixing fraction $f$ increases. However, this effect is not observed in periodic SLs for non-Bragg-reflected phonons, such as $\lambda_{1}$ and $\lambda_{2}$ (Fig.~\ref{fig:transmission_SL}), as their transmission is already near unity due to the strong mode-conversion facilitated by the well-defined dispersion relation \cite{maranets2024phonon}. In such cases, the primary effect of interface mixing---disruption of mode-conversion---prevails, resulting in a reduction in transmission.

It is noteworthy that interface scattering induces interference, constructive or destructive, based on the layering pattern, which is analogous to the well-established behavior of light waves (or, more generally, photon waves). This principle forms the basis of hypotheses proposed in earlier studies on periodic and aperiodic SLs \cite{luckyanova2012coherent,wang2014decomposition,maldovan2015phonon,qiu2015effects}. Notably, for aperiodic SLs, it has been theorized that the localization of coherent phonons, arising from the aperiodic arrangement of interfaces, is weakened by interface scattering introduced through mixing \cite{qiu2015effects}. This hypothesis has been used to explain the higher thermal conductivity observed in aperiodic SLs with mixed interfaces compared to those with perfect interfaces. In contrast, the present work offers a fundamentally different explanation: interface mixing facilitates significant coherent mode-conversion for low-wavevector phonons, thereby creating additional heat conduction pathways and enhancing the overall thermal conductivity of the aperiodic SL.

\begin{figure}
    \centering
    \includegraphics[width=0.9\textwidth]{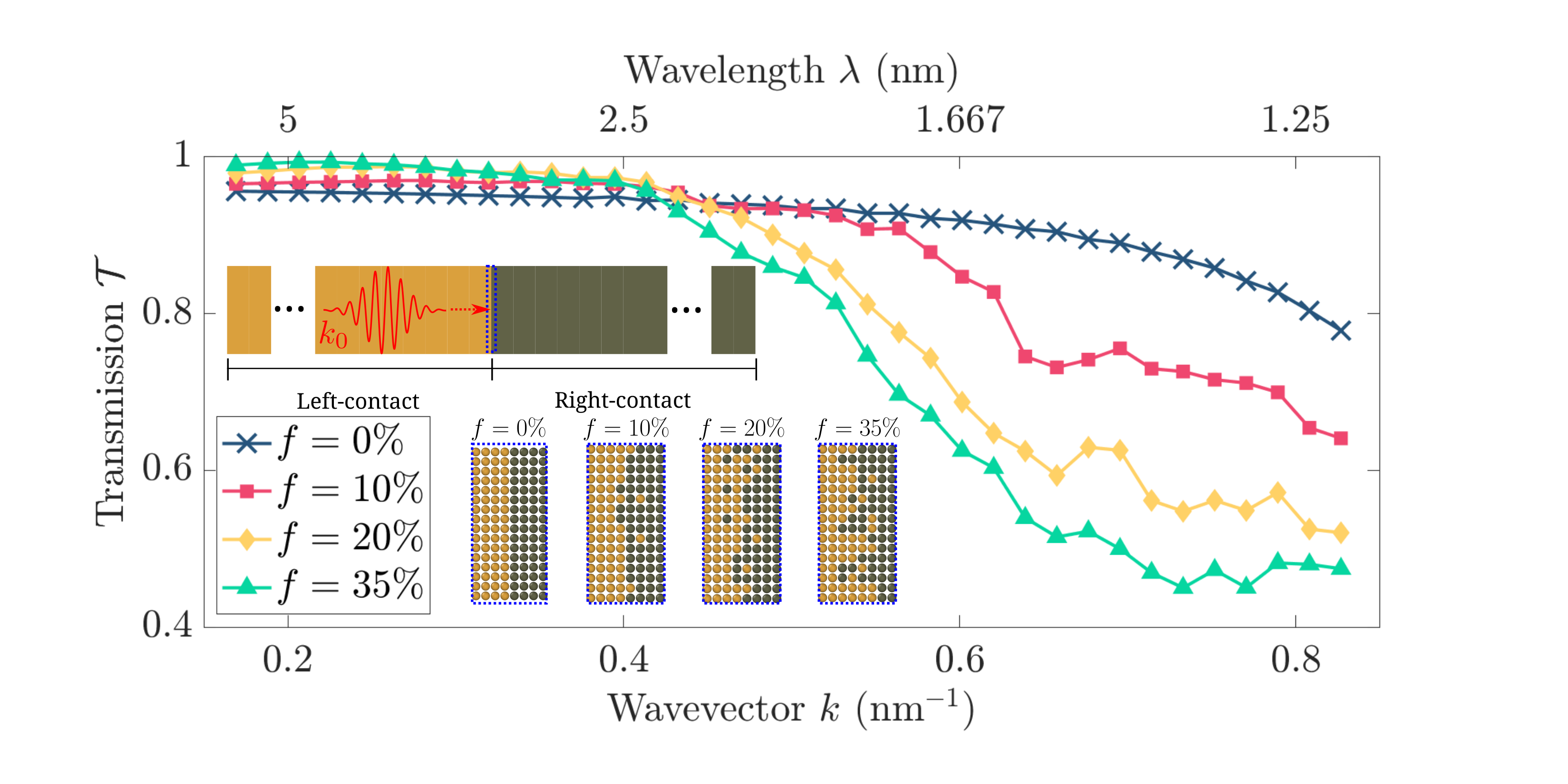}
    \caption{Transmission $\mathcal{T}$ versus wavevector $k$ (or inverse of wavelength $\lambda$) for the LA-mode incoherent phonon wave-packet propagating across a single m40-m90 interface of differing mixing fractions $f$ as shown in the schematic illustration inset. $\mathcal{T}$ is evaluated with Eqn.~\ref{eqn:transmission_calc}.}
    \label{fig:transmission_SingleInterface}
\end{figure}

Finally, it is critical to distinguish the contrasting effects of interface mixing on low-wavevector (long-wavelength) and high-wavevector (short-wavelength) phonons. High-wavevector phonons, characterized by shorter wavelengths, are strongly scattered by mixed interfaces, similar to the behavior observed in systems with point defects or alloys \cite{garg2011role,gurunathan2020analytical}. This pronounced scattering attenuates coherent mode-conversion, resulting in a broadband suppression of transmission, as evidenced for mid- and high-wavevector phonons in both periodic (Fig.~\ref{fig:transmission_SL}) and aperiodic SLs (Fig.~\ref{fig:transmission_RML}). In contrast, low-wavevector phonons, with their longer wavelengths, are less susceptible to scattering and destructive interference, allowing them to retain a significant degree of coherent mode-conversion. This phenomenon accounts for the substantial transmission observed for low-wavevector phonons in interface-mixed SLs, as demonstrated in Figs.~\ref{fig:transmission_SL} and \ref{fig:transmission_RML}. The transmission spectra for a single m40-m90 interface (Fig.~\ref{fig:transmission_SingleInterface}) further substantiates these observations, highlighting the pronounced impact of interface mixing on high-wavevector phonons and its comparatively weaker effect on low-wavevector phonons. For example, as the mixing fraction $f$ increases, the single-interface transmission of a phonon mode with a wavevector of 0.6~nm$^{-1}$ decreases significantly, from nearly 0.9 across the perfect interface ($f = 0$) to only 0.6 across the interface with $f = 35\%$. This substantial reduction in single-interface transmission can profoundly diminish the cumulative transmission across 128 interfaces studied in this work, effectively resulting in near-zero transmission for the entire interface-mixed SL structure. For low-wavevector phonons, the single-interface transmission partially increases with $f$. The lack of significant suppression of transmission facilitates the low-wavevector phonons to mode-convert and thus transmit prominently across the interface-mixed SLs unlike the higher wavevectors.

\section{Conclusion}
In summary, this study explores the influence of interface mixing on coherent heat conduction in periodic and aperiodic SLs. Using atomistic wave-packet simulations, we reveal that interface mixing generally suppresses the transmission of low-wavevector phonons in periodic SLs while enhancing it in aperiodic SLs. This behavior arises from two competing effects of interface mixing on coherent mode-conversion: 
\begin{enumerate}
    \item Mixed interfaces scatter incident incoherent phonons originating from the contact or heat bath, disrupting the formation of coherent phonons through constructive interference induced by the layering pattern.
    \item The scattering of incoherent phonons by mixed interfaces introduces new opportunities for constructive interference, potentially generating additional coherent phonon transport pathways across the SL.
\end{enumerate}
The first effect reduces transmission, while the second effect enhances it. The second effect significantly boosts transmission for a wide range of wavevectors in aperiodic SLs and for Bragg-reflected modes in periodic SLs by leveraging the created coherent phonon channels to promote transmission. In contrast, in periodic SLs, the second effect is negligible for non-Bragg-reflected modes, as their transmission is already near unity due to substantial coherent mode-conversion; in these cases, the first mechanism dominates, leading to reduced transmission.

For intermediate- and high-wavevector phonons, transmission is suppressed by interface mixing in both periodic and aperiodic SLs. This occurs because shorter-wavelength phonons are more susceptible to interface scattering, which reduces their transmission at individual interfaces and results in near-zero transmission across SL devices with numerous interfaces.

Our findings provide critical insights into the role of interface imperfections on phonon coherence and heat conduction, particularly in the context of real-world applications where structural irregularities are practically unavoidable. This work advances the understanding of phonon transport mechanisms in SLs and highlights the intricate interplay between interface mixing and coherent phonon behavior.

\begin{acknowledgments}
The authors gratefully acknowledge the financial support from the National Science Foundation (CBET-2047109). Maranets thanks the support from the Nevada NASA Space Grant Graduate Research Opportunity Fellowship and the Nuclear Power Graduate Fellowship from the Nuclear Regulatory Commission. Doe thanks the support from the Nevada NASA Space Grant Undergraduate Scholarship. Additionally, the authors would like to acknowledge the support provided by the Research and Innovation team and the Cyberinfrastructure Team in the Office of Information Technology at the University of Nevada, Reno, for facilitating access to the Pronghorn High-Performance Computing Cluster.
\end{acknowledgments}

\section*{Author Contributions}
\textbf{Theodore Maranets:} Writing – review \& editing (equal), Writing – original draft (equal), Software (lead), Methodology (lead), Formal analysis (equal), Conceptualization (equal). 
\textbf{Evan Doe:} Writing – review \& editing (equal), Writing – original draft (equal), Software (equal), Methodology (equal), Formal analysis (equal), Conceptualization (equal). \textbf{Yan Wang:} Writing – review \& editing (equal), Writing – original draft (equal), Supervision (lead), Funding acquisition (lead), Formal analysis (equal), Conceptualization (equal).

\section*{Data Availability}
The data that support the findings of this study are available from the corresponding author upon reasonable request.

\bibliography{references}

\end{document}